\documentclass[conference]{IEEEtran}
\IEEEoverridecommandlockouts
\usepackage{cite}
\usepackage{amsmath,amssymb,amsfonts}
\usepackage{algorithmic}
\usepackage{graphicx}
\usepackage{textcomp}
\usepackage{xcolor}
\usepackage{lmodern,fancyhdr,secdot,float,amsmath,amsfonts,amsthm,amssymb}
\usepackage{multirow}
\usepackage{array}

\def\BibTeX{{\rm B\kern-.05em{\sc i\kern-.025em b}\kern-.08em
    T\kern-.1667em\lower.7ex\hbox{E}\kern-.125emX}}
\begin{document}

\title{RIS-measurements for Codebook Design \\

\thanks{The presented work has been funded by the National Science Centre in Poland within the project (no. 2021/43/B/ST7/01365) of the OPUS program}
}

\author{\IEEEauthorblockN{1\textsuperscript{st} Pawel Hatka}
\IEEEauthorblockA{\textit{Poznan University of Technology} \\
Poznan, Poland \\
pawel.hatka@student.put.poznan.pl}
\and
\IEEEauthorblockN{2\textsuperscript{nd} Marcel Garczyk}
\IEEEauthorblockA{\textit{Poznan University of Technology} \\
Poznan, Poland  \\
marcel.garczyk@student.put.poznan.pl}
\and
\IEEEauthorblockN{3\textsuperscript{rd} Pawel Placzkiewicz}
\IEEEauthorblockA{\textit{Poznan University of Technology} \\
Poznan, Poland  \\
pawel.placzkiewicz@student.put.poznan.pl}
\and
\IEEEauthorblockN{4\textsuperscript{th} Dawid Brzakala}
\IEEEauthorblockA{\textit{Poznan University of Technology} \\
Poznan, Poland \\
dawid.brzakala@student.put.poznan.pl}
\and
\IEEEauthorblockN{5\textsuperscript{th} Krzysztof Cichon}
\IEEEauthorblockA{\textit{Poznan University of Technology} \\
Poznan, Poland \\
krzysztof.cichon@put.poznan.pl}
\and
\IEEEauthorblockN{6\textsuperscript{th} Adrian Kliks }
\IEEEauthorblockA{\textit{Poznan University of Technology} \\
Poznan, Poland \\
adrian.kliks@put.poznan.pl}
}

\maketitle

\begin{abstract}
Reconfigurable Intelligent Surfaces (RIS) have gained significant attention for some time. Thanks to the possibility of individual steering of each reflecting element of the boards, they are envisaged to impact the propagation environment significantly. In this work, we concentrate on the practical verification of this concept. We present the results of detailed measurements of the reflection characteristics of the RIS boards, which have been conducted intentionally in the real environment. Various potential impacting factors have been considered (impact of azimuth and elevation angle, polarization, number of RIS boards, and distance). Achieved measurement results constituted the basis for conceptual analysis on the practical possibility of creating a codebook (consisting of RIS patterns - codewords) for some applications. 
\end{abstract}
\footnote{Copyright © 2024 IEEE. Personal use us permitted. For any other purposes, permission must be obtained from the IEEE by emailing pubspermissions@ieee.org. This is the author’s version of an article that has been published in the proceedings of  2024 20th International Conference on Wireless and Mobile Computing, Networking and Communications and published by IEEE. Changes were made to this version by the publisher prior to publication, the final version of record is available at: https://ieeexplore.ieee.org/document/10770406}
\begin{IEEEkeywords}
RIS, codebook design, measurements, reflection characteristics
\end{IEEEkeywords}

\section{Introduction}
The Reconfigurable Intelligent Surfaces (RISes) are treated as one of the most promising technological solutions for future wireless networks \cite{Di2022,ETSI1}. These (semi)-passive arrays consist of many antenna elements (from just a few to thousands), which are typically densely packed in a specific area. Each of these elements may be individually controlled, impacting the behavior of the whole board. Depending on the manufacturing technology, these singular elements may, for example, impact the phase shift of the incident signal, absorb the signal, or pass through the whole or part of the signal \cite{Joy2024,Gong2024}. Consequently, RIS may manipulate the propagation environment through a dynamic change of the reflection pattern of the board. As RIS control may be done remotely, the boards allow the modification of radio transmission characteristics to achieve desired propagation effects \cite{Renzo2019}. Various interesting applications can be identified while analyzing vast literature in that domain, such as tracking of moving users, improvement of sensing capabilities, interference minimization, etc. \cite{Liu2022,Shi2024}.

As mentioned, RIS are typically treated as quasi-passive elements of the network infrastructure, where the antenna elements themselves are passive without power amplification possibilities \cite{ETSI1}, making this solution energy-effective. Moreover, the performance of RIS depends on the number of these elements, which directly impacts the RISes ability to influence the incident signal. Each element can be set individually to operate on one of $M$ available modes; thus, for $N$-element RIS, there is a huge space of $M^N$ different patterns. There is a need for an efficient definition of the rules for codebook design, taking into account various practical limitations of the particular board manufactured with a specified technology.

In this work, we discuss the possibility of codebook design for the  RIS manufactured in the OpenSourceRIS project \cite{GitHub1,Heinrichs2023}. We present the results of laboratory measurements of the reflection characteristics of these boards located in a natural (non-ideal) environment. Our intention was to verify, by measurements, if there is room for recreation of the codebook consisting of specific RIS patterns, which could be used in specific application scenarios. 

`

\section{Description of the Measurement Scenario}
The article considers three measurement scenarios. The first involved measuring the three-dimensional reflective characteristics of RIS at various distances from the transmitter and receiver (denoted hereafter as Scenario 1). The impact of the RIS operation was also compared to the reflective characteristics of a copper plate with dimensions equal to those of the RIS matrix. Scenario 2 focuses on the impact of the polarization, and the last - Scenario 3 - analyzed the effect of combining two RIS boards together to create a bigger surface.
\subsection{Overall setup}
\begin{figure}[!hbt]
\centering
\includegraphics[width=1\linewidth,height=5.3cm]{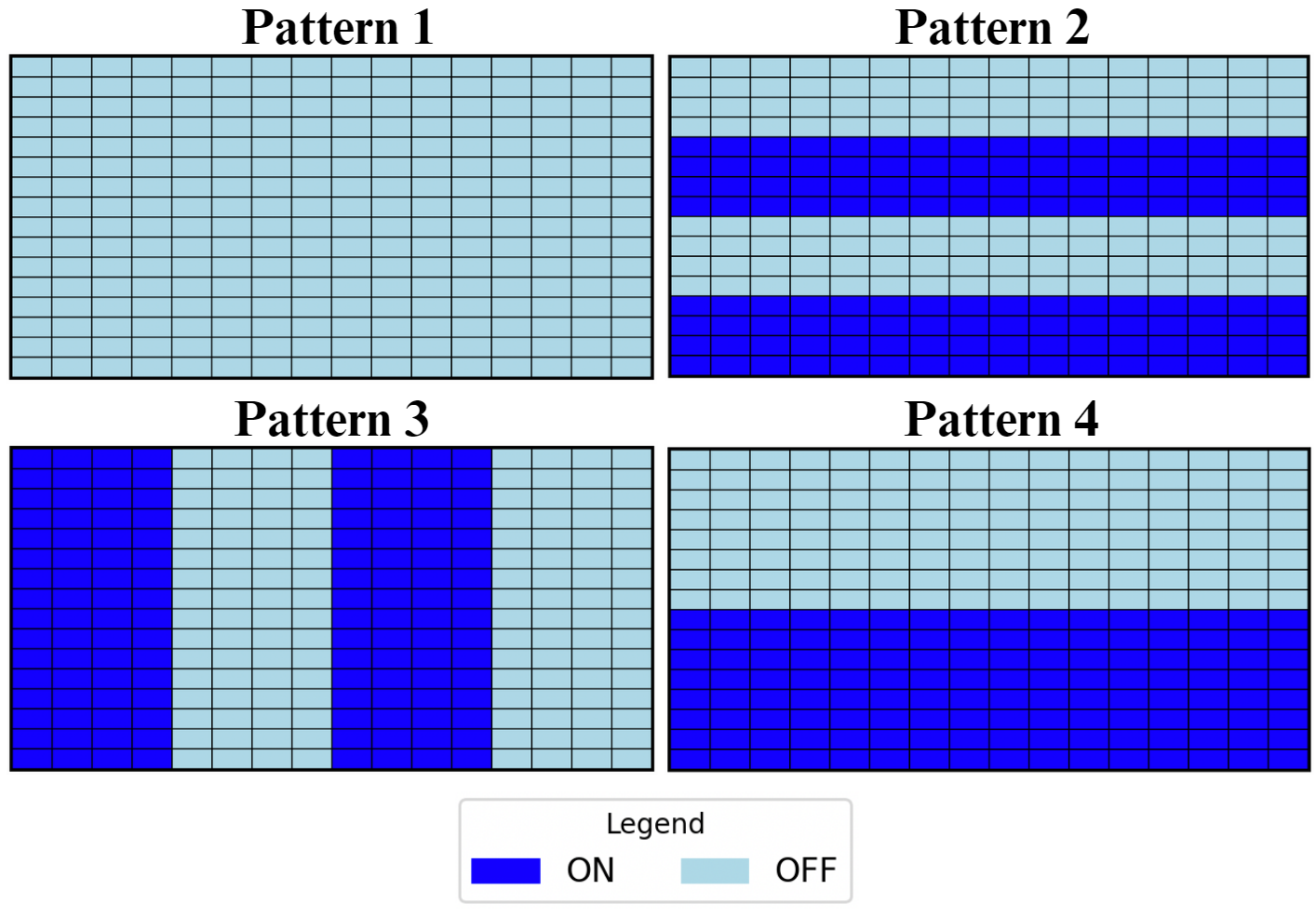}
\caption{View of the measured patterns.}
\label{all_patterns}
\end{figure}

The following measurement setup was assembled in the laboratory to conduct these studies. A signal with a frequency of 5.5 GHz and power of -10~dBm was generated using a Rohde \& Schwarz SMM100A generator and transmitted towards the RIS array using a METIS 5.1-5.8~GHz directional antenna with vertical polarization. On the receiving side, a Rohde \& Schwarz FSV3000 spectrum analyzer connected to an identical antenna with the same polarization was used. The individual settings of the analyzer are presented in Table \ref{Settings}. The transmitting and receiving antennas were directed at the RIS array, making it possible to measure the level of reflected power observed at the receiver. The RIS matrix has $16 \times 16$ reflective elements that are individually controlled by one bit, meaning each element can either reflect the incoming signal without intentionally changing its phase or change it by 180 degrees. It is important to note that even this simple variability creates a very wide range of possible configurations, i.e., $2^{256}$. 
\begin{table}[H]
\caption{Parameters of Spectrum Analyzer and Signal Generator  }
\begin{center}
\begin{tabular}{|>{\centering\arraybackslash}m{3cm}|c|c|c|}
\hline
\textbf{Device} & \textbf{Parameter} & \textbf{Value} & \textbf{Unit} \\
\hline
\multirow{7}{*}{\centering\textbf{Spectrum Analyzer}} & \text{Span} & \text{0} & \text{Hz} \\
\cline{2-4}
 & \text{RBW} & \text{500} & \text{Hz} \\
\cline{2-4}
 & \text{Sweep} & \text{50} & \text{ms} \\
\cline{2-4}
 & \text{Ref. level} & \text{-30} & \text{dBm} \\
\cline{2-4}
 & \text{Center frequency} & \text{5.5} & \text{GHz} \\
\cline{2-4}
 & \text{Noise level} & \text{-110} & \text{dBm} \\
\cline{2-4}
 & \text{Mode} & \multicolumn{2}{c|}{\text{MaxHold}} \\
\hline
\multirow{2}{*}{\centering\textbf{Signal Generator}} & \text{Transmitted power} & \text{-10} & \text{dBm} \\
\cline{2-4}
 & \text{Mode} & \multicolumn{2}{c|}{\text{CW}} \\
\hline
\end{tabular}
\label{Settings}
\end{center}
\end{table}
The entire measurement process was automated using software written in Python, which allowed remote control of the spectrum analyzer and signal generator, rotating the RIS surface in both elevation and azimuth angles and changing the pattern set on the RIS. The mentioned pattern defines which of the 256 elements of the matrix change phase by 180 degrees, i.e., which are active and which are not. Due to complexity issues in our study, these 256 elements were divided into 64 groups of four elements each; hence, there are four such groups in one row. Controlling the matrix involves selecting which of the four elements in one group are to be active. Group numbering starts from index 0, and the first group is located in the upper left corner of the matrix. The selection is made by sending a sequence of numbers in hexadecimal code to the RIS, where the position in the sequence determines the group. Thus, 0 - 0000 means all elements of the group are off, F - 1111 means all are on, and 3 - 0011 means the last two are on, etc. An example sequence:
!0X800...000 - means that only the first element of group 0 will be activated \cite{GitHub1}. Measurements were conducted for 27 arbitrarily selected patterns, but due to article space limitations, results for patterns shown in Fig.\ref{all_patterns} will be presented.

\subsection{3D characteristic}
\begin{figure}[H]
\centering
\includegraphics[width=1\linewidth, height = 6cm]{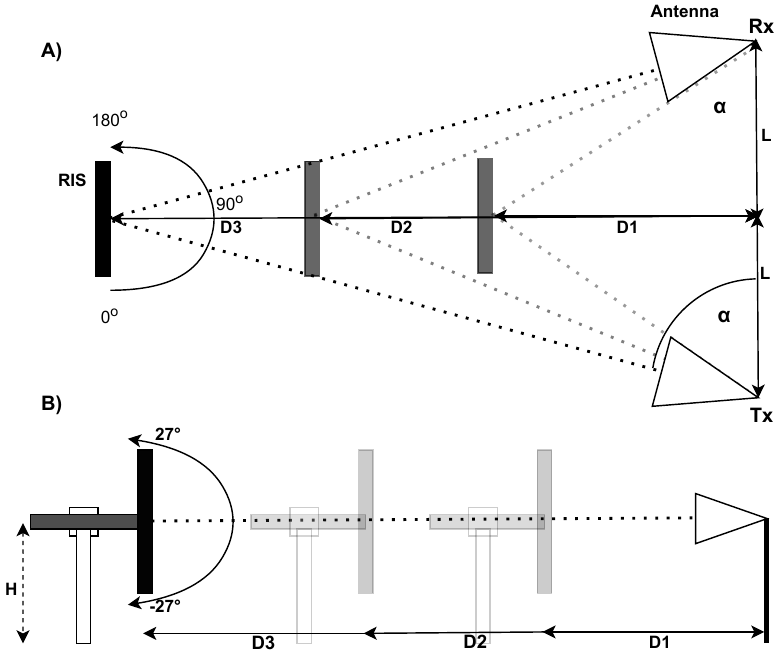}
\caption{Measurement scheme: A - top view, B - side view}
\label{3D_schemat}
\end{figure}

To perform the spatial measurement of the reflective characteristics, it was necessary to ensure the capability of rotating the matrix in both horizontal and vertical planes. During the experiment, the change in azimuth and elevation angles was facilitated by a specially designed and manufactured arm equipped with a motor and control system. Reflective characteristics measurements were carried out at three different distances, as illustrated in the measurement diagram (Fig.\ref{3D_schemat}).
In the first case, the distance between the antenna line and the RIS matrix line was $D1=1 m$. Then, this distance was increased to $D2=1.5 m$  and $D3=2 m$. In each case, the distance between the antennas was 2 meters ($L=1m$), resulting in an antenna orientation angle towards the RIS of 45.0$^\circ$ for 1 meter, 56.3$^\circ$ for 1.5 meters, and 63.4$^\circ$ for 2 meters. To obtain accurate results, the head with the mounted matrix rotated in the azimuth plane with a step of 1.8$^\circ$, starting from an angle of 0$^\circ$ and ending at 180$^\circ$. At 90$^\circ$, the matrix plane was oriented parallel to the antenna line. The antennas and matrix were placed at the height of $H=1.3 m$.
\subsection{Vertical polarization}
The second scenario involved performing the same measurement using identical patterns set on the RIS but with the board positioned vertically (i.e., rotated by 90$^\circ$), with the shorter side of the array facing upwards. This measurement was conducted for a distance of 1.5 meters.
\subsection{Additional RIS}
\begin{figure}[h]
\centering
\includegraphics[width=0.8\linewidth,height=2.5cm]{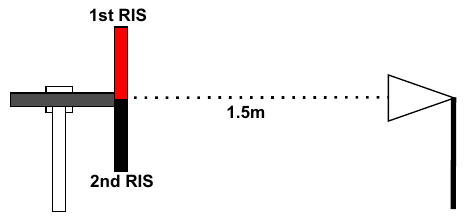}
\caption{Side view after additional RIS was added.}
\label{2_RIS_schemat}
\end{figure}

The next measurement scenario considered was the addition of a second array to the measurement setup as shown on (Fig. \ref{2_RIS_schemat}), i.e., the second RIS board was bonded with the first one, doubling the reflection area and number of available elements. The RIS was positioned horizontally. For this case, the measurement was conducted at a distance of 1.5 meters only for a zero elevation angle due to the construction limitations of the rotating head. The addition of the second array required raising the antennas to a height of 1.44 meters, so that the center of the antennas was at the height of the junction of the two plates. The remaining parameters stayed the same.

\section{Measurement Results}
Results obtained in scenarios 1 and 2 are presented as heatmap plots. They depict the level of received power by the receiving antenna according to the diagram (Fig. \ref{2_RIS_schemat}), as a function of the matrix rotation in both the azimuth and elevation planes. The Y-axis represents the rotation of the matrix in the elevation plane, while the X-axis represents the rotation in the azimuth direction relative to the antenna line.

\subsection{Scenario 1 - 3D Characteristic}
\begin{figure}[!b]
    \centering
    \includegraphics[width=1\linewidth,height=5.3cm]{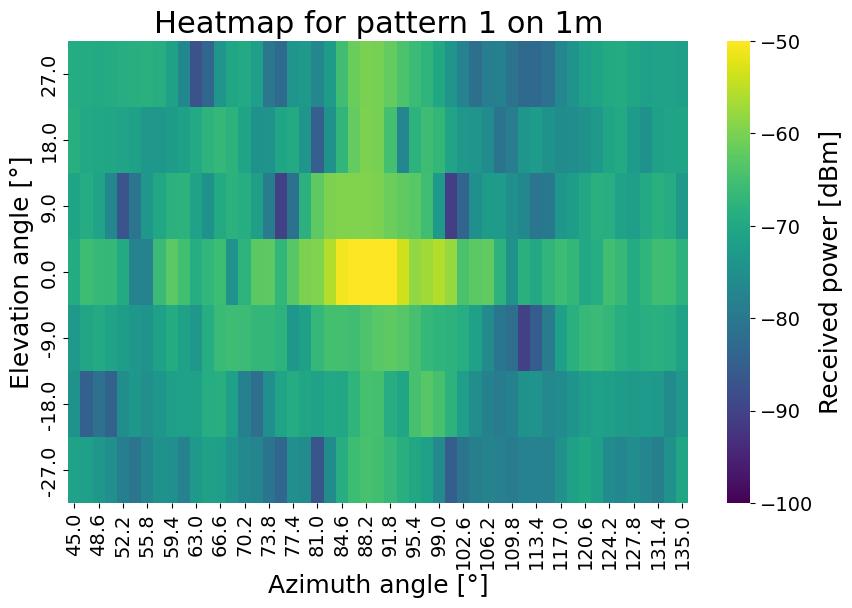}
    \includegraphics[width=1\linewidth,height=5.3cm]{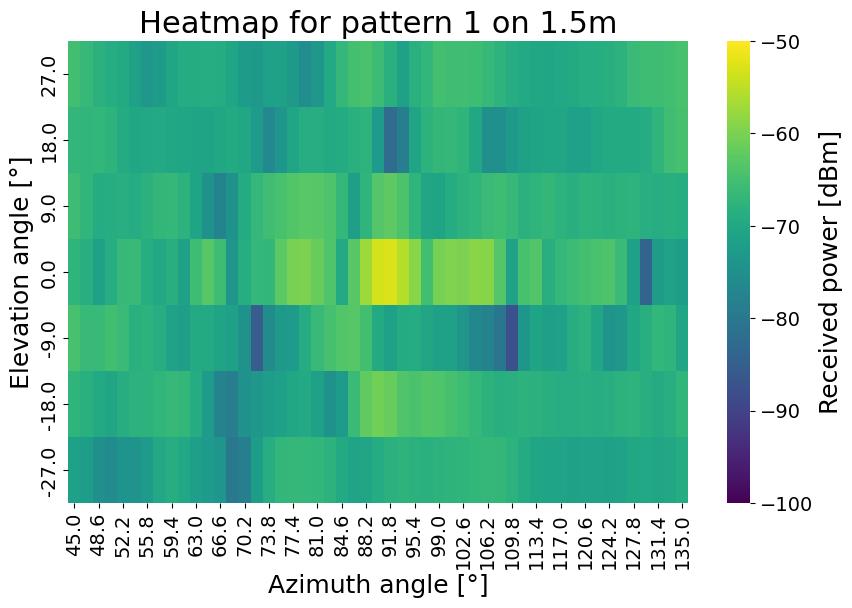}
    \includegraphics[width=1\linewidth,height=5.3cm]{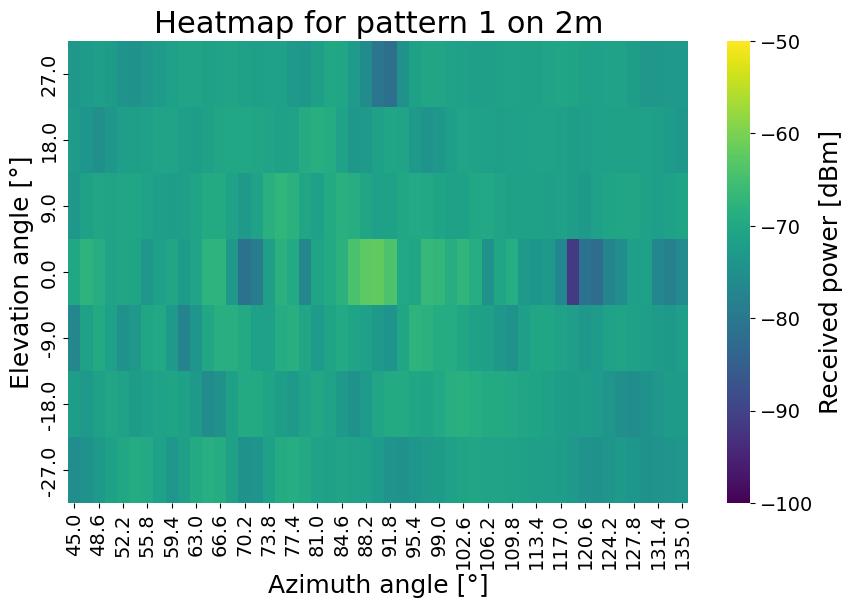}
    \caption{Results for 1st pattern.}
    \label{1st pattern}
\end{figure}

For the first pattern (Fig. \ref{1st pattern}), where all elements of the array are turned off, we obtain the maximum received power at a matrix angle of 90°, which is when the surface is oriented parallel to the antenna line—where the angle of reflection equals the angle of incidence.

\begin{figure}[!htb]
    \centering
\includegraphics[width=1\linewidth,height=5.1cm]{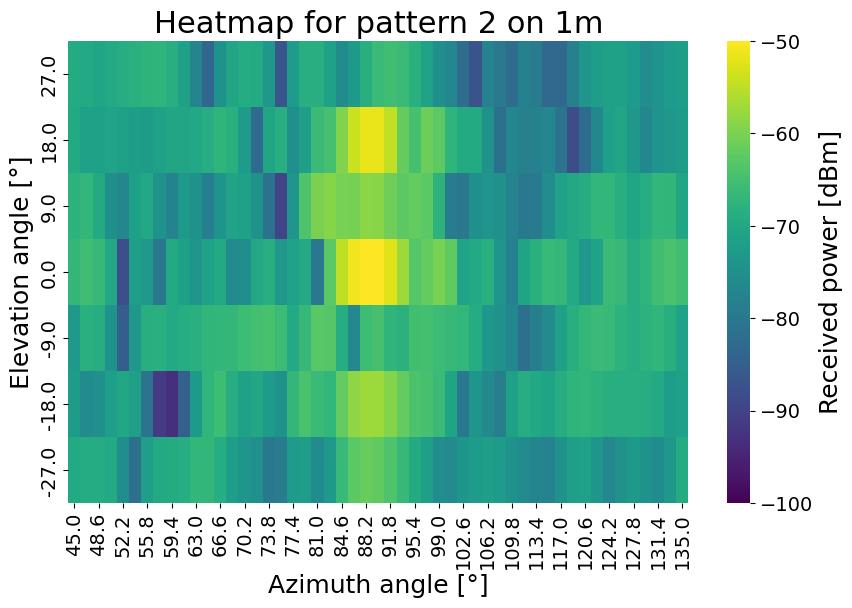}
\includegraphics[width=1\linewidth,height=5.1cm]{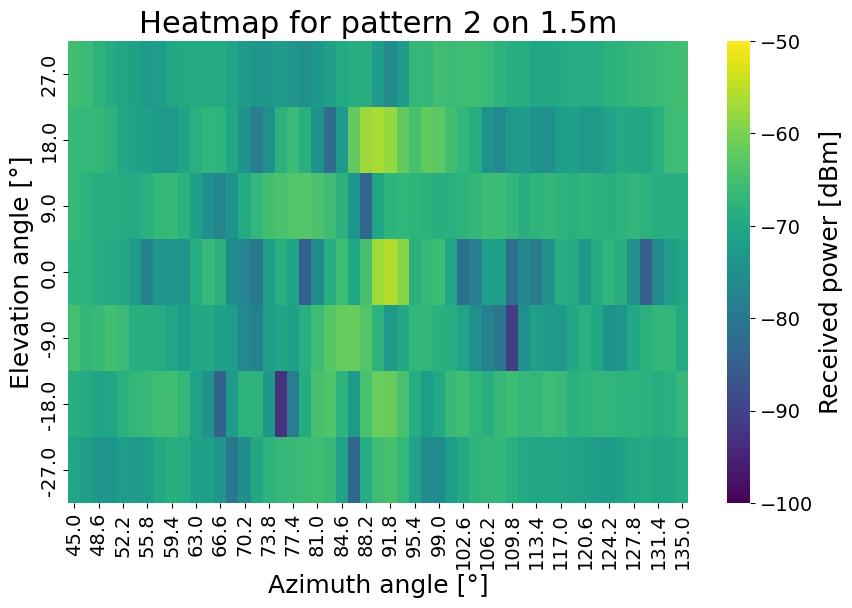}
\includegraphics[width=1\linewidth,height=5.1cm]{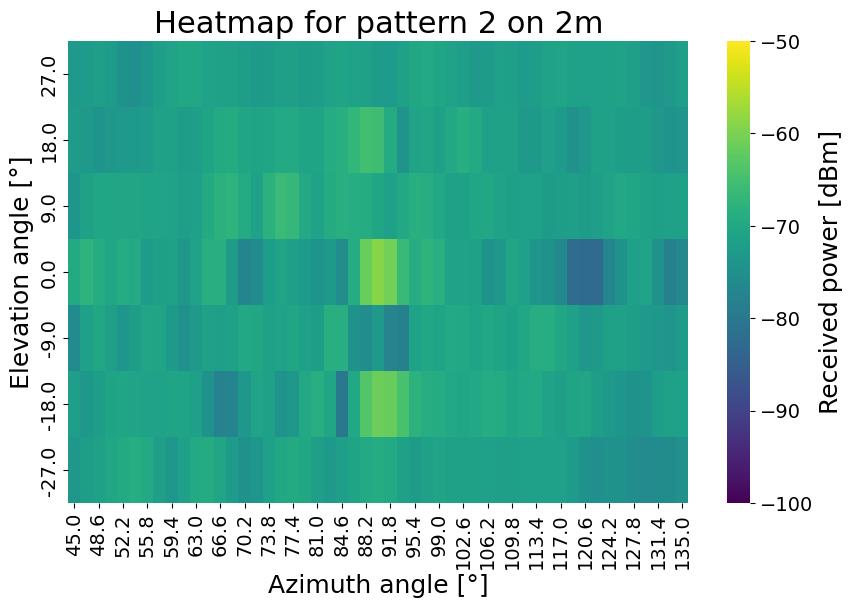}
    \caption{Results for 2nd pattern.}
    \label{2nd pattern}
\end{figure}

The results for patterns 2 (Fig. \ref{2nd pattern}), which consist of 4 rows wide horizontal stripes, reveal an interesting phenomenon. We observe a stronger peak in power apart from 0° at the angle of 18° elevation, also a smaller peak of power can be observed at -18° elevation. As the distance increases to 2 meters, the position of the stronger peak shifts around the 0° elevation angle, which requires further research.

\begin{figure}[!htb]
    \centering
\includegraphics[width=1\linewidth,height=5.1cm]{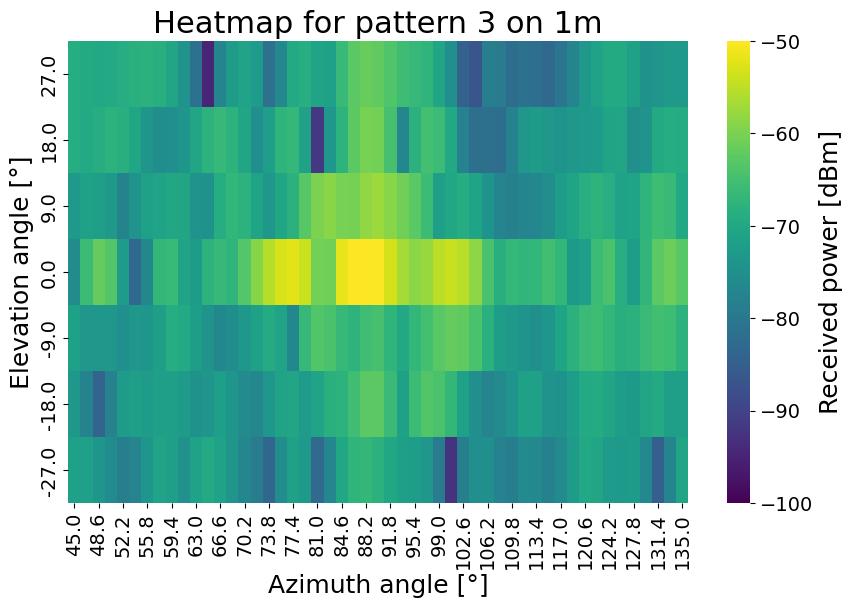}
\includegraphics[width=1\linewidth,height=5.1cm]{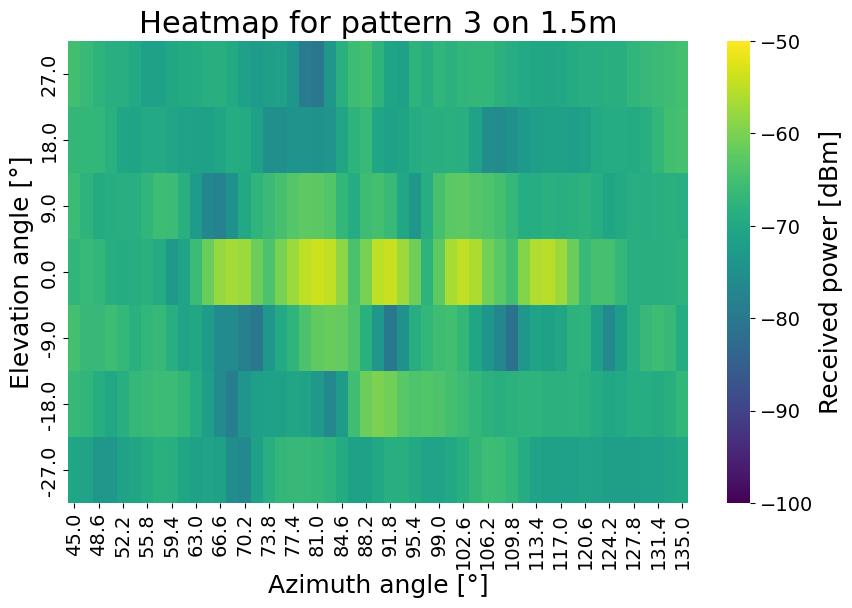}
\includegraphics[width=1\linewidth,height=5.1cm]{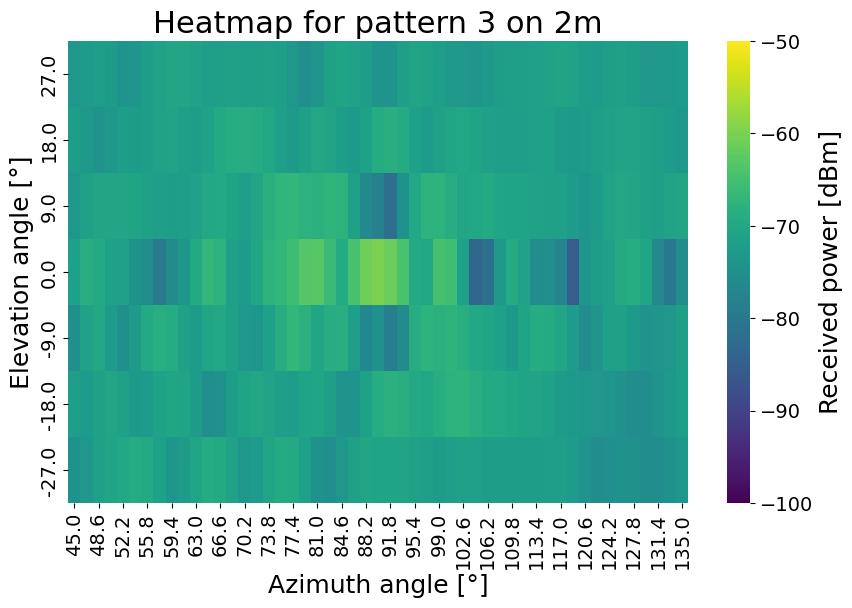}
    \caption{Results for 3rd pattern.}
    \label{3rd pattern}
\end{figure}

In the graph (Fig. \ref{3rd pattern}) for the third pattern, which consists of stripes 4 columns wide, an interesting effect can be observed, similar to the effect occurring for the second pattern. For an elevation angle of 0°, we obtain several peaks in received power as a function of the azimuth angle, particularly noticeable for a distance of 1.5 meters, although also observable for 1 meter. Between these peaks, the level of received power is significantly lower. For a distance of 2 meters, this effect disappears. This effect is most likely caused by the fact that, as the matrix rotates, the beam sometimes reflects off the activated part of the matrix and, at other times, off the deactivated part. 

\begin{figure}[!htb]
    \centering
\includegraphics[width=1\linewidth,height=5.1cm]{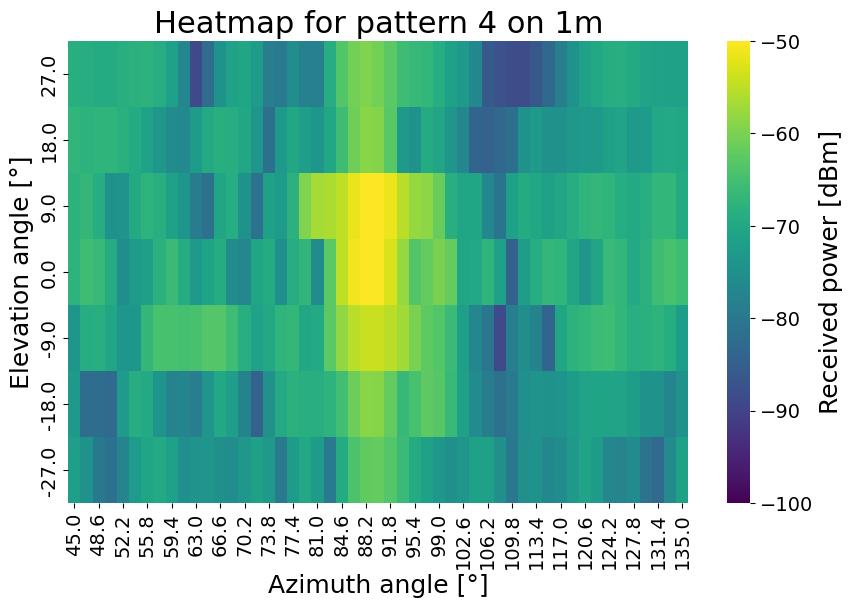}
\includegraphics[width=1\linewidth,height=5.1cm]{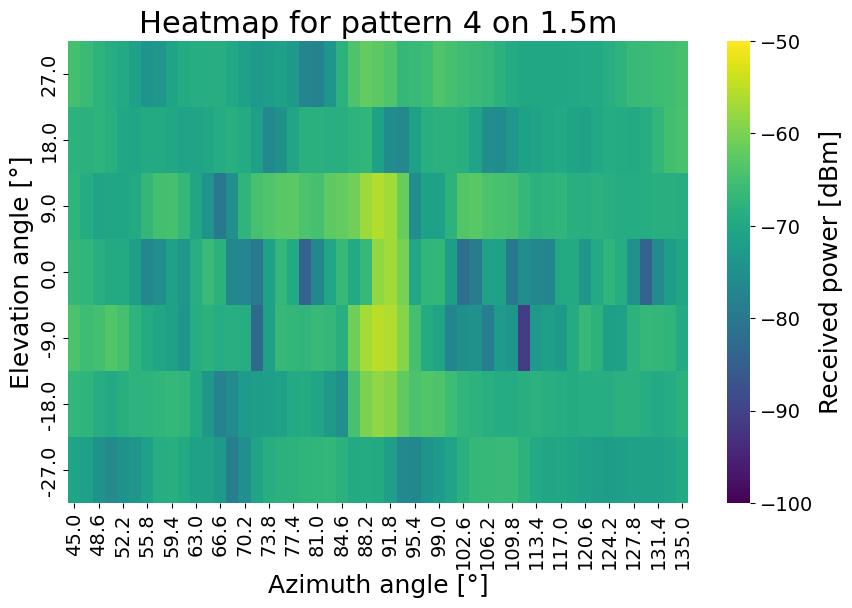}
\includegraphics[width=1\linewidth,height=5.1cm]{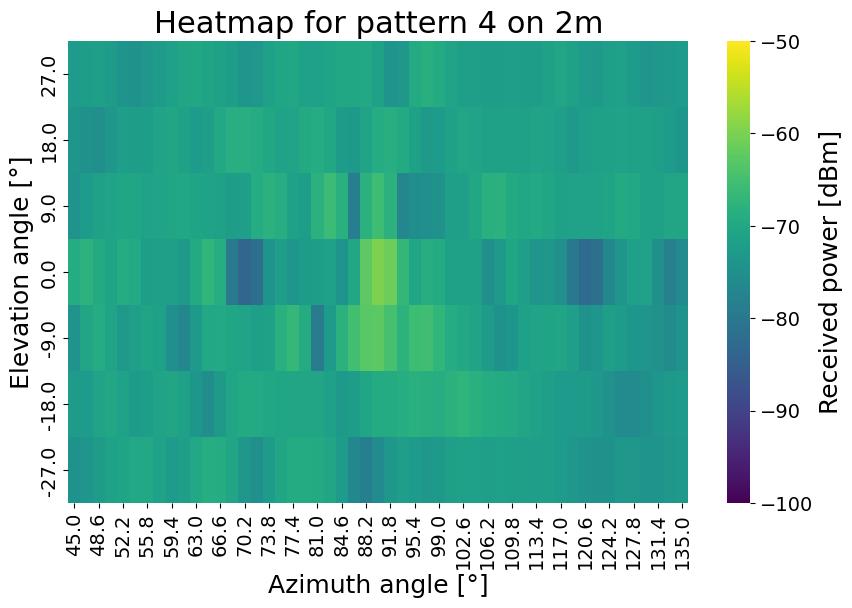}
    \caption{Results for 4th pattern.}
    \label{4th pattern}
\end{figure}

For the fourth pattern (Fig. \ref{4th pattern}), the effect of changing the pattern is most noticeable for a distance of 1 meter and to a slightly lesser extent for 1.5 meters. Activating the lower half of the RIS caused a distribution of part of the received power at elevations of -9° and 9°. This is another pattern showing that the matrix setting can influence how the signal is received. Changing the distance resulted in a decrease in received power at these elevations.

\begin{figure}[!htb]
    \centering
\includegraphics[width=1\linewidth,height=5.1cm]{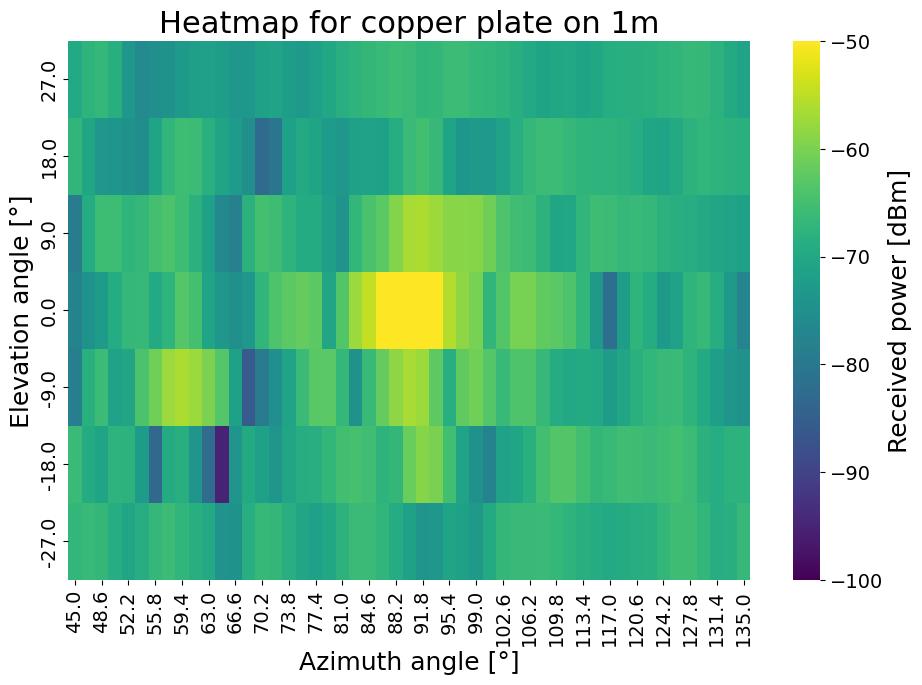}
\includegraphics[width=1\linewidth,height=5.1cm]{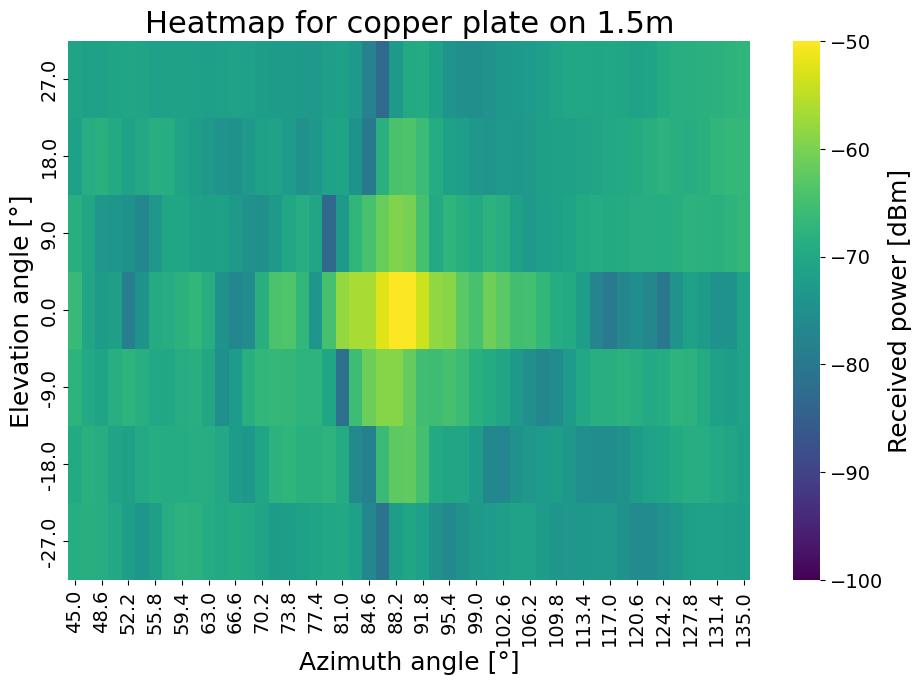}
\includegraphics[width=1\linewidth,height=5.1cm]{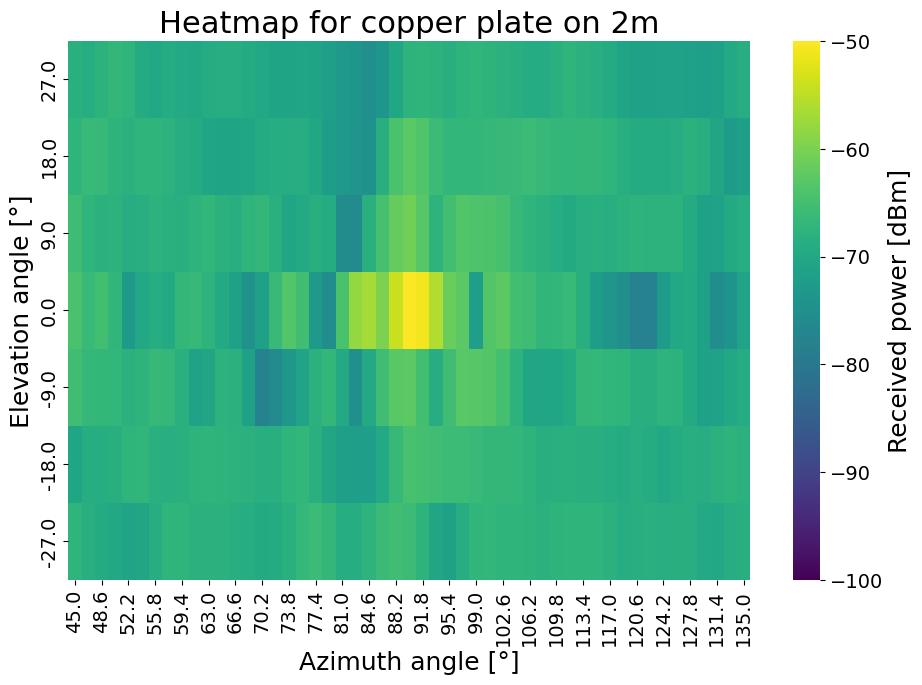}
\caption{Results for copper plate.}
\label{copper_plate}
\end{figure}

To verify the results of the individual patterns and ensure that some symmetries, whether in elevation or azimuth, are not coincidental, the antenna array was replaced with an appropriately cut fragment of copper plate. The results (Fig. \ref{copper_plate}) are similar to the first scenario—when all elements of the array were turned off. Again, the maximum received power is obtained at an angle of 90°, with the array positioned parallel to the antenna line. This verification confirms that changing the pattern can significantly affect the quality of the received signal in a given environment.

\subsection{Second scenario}
\begin{figure}[!t]
\centering
\includegraphics[width=1\linewidth,height=5.1cm]{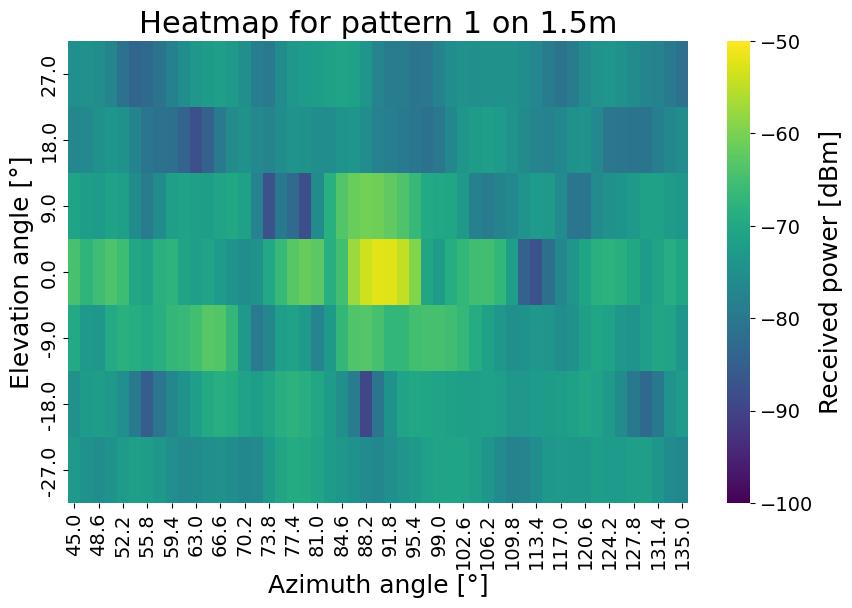}
\includegraphics[width=1\linewidth,height=5.1cm]{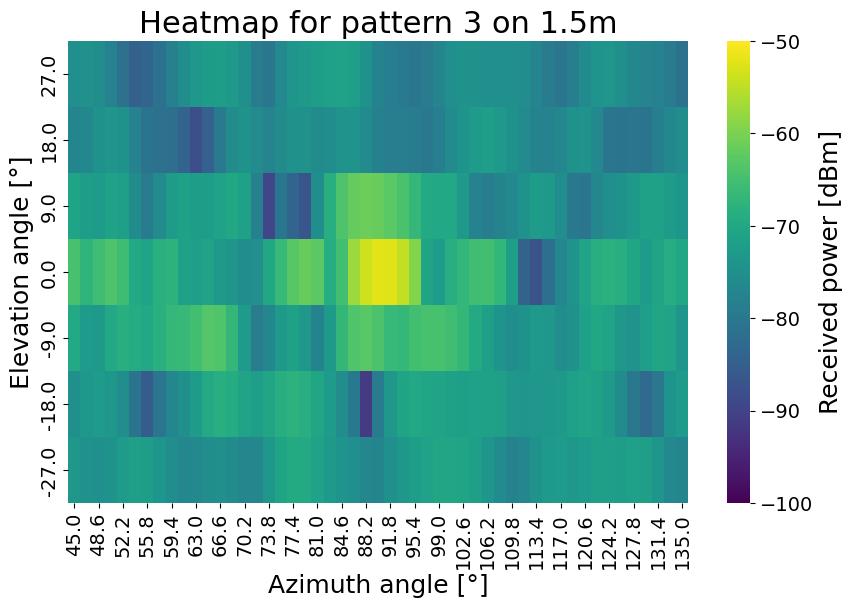}
\caption{Results for second scenario.}
\label{Second_scenario}
\end{figure}

Changing the orientation of the antenna array from horizontal to vertical yielded the expected results in the form of no impact of the pattern change on the received power. The RIS consists of a system of linearly polarized elements, so changing the polarization of the array without simultaneously changing the polarization of the transmitting and receiving antennas results in the disappearance of the pattern's influence on the received power, as can be observed in the presented graphs (Fig. \ref{Second_scenario}). The received power distribution was largely the same for all 27 tested patterns.

\subsection{Third scenario}

\begin{figure}[!t]
    \centering
\includegraphics[width=1\linewidth,height=5cm]{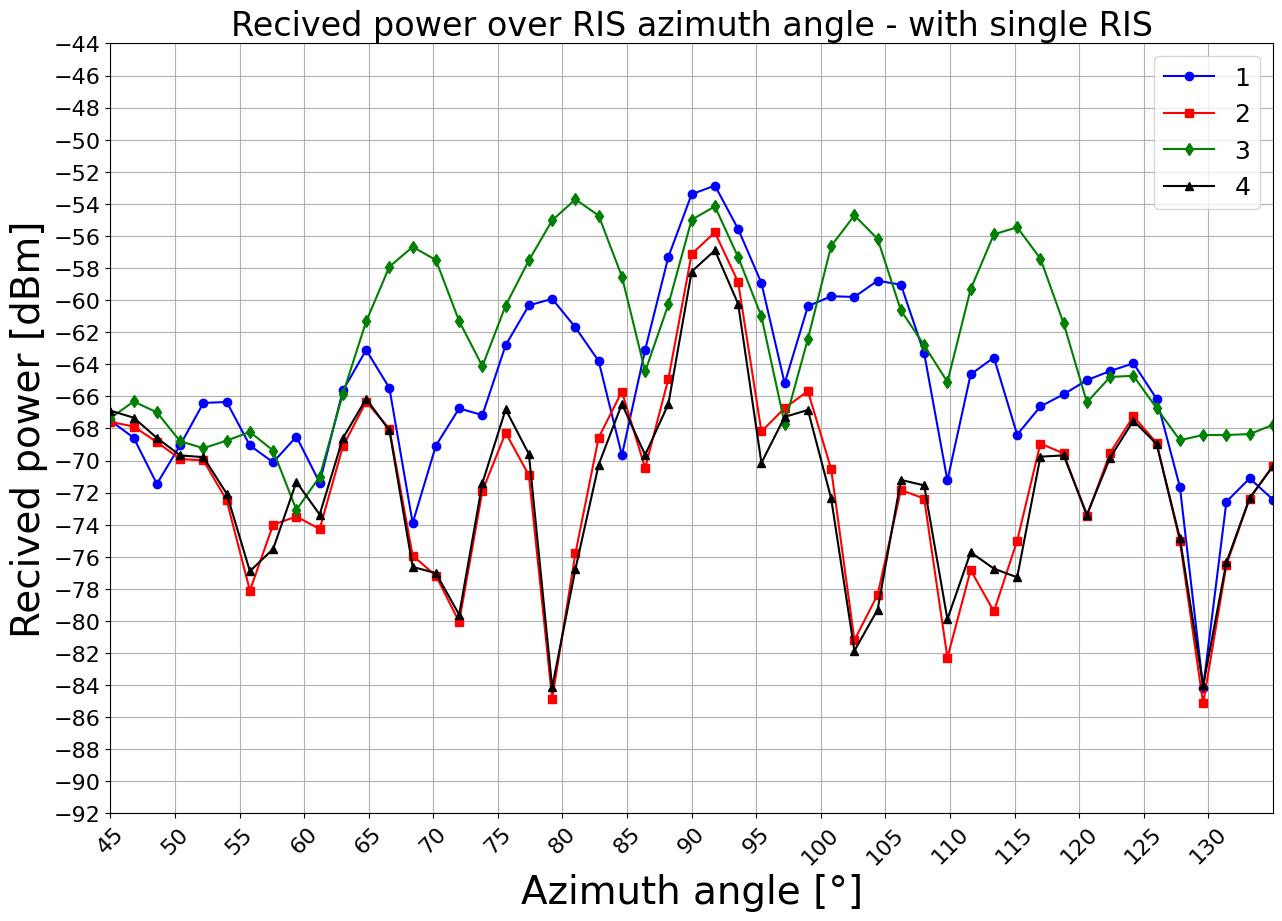}
\includegraphics[width=1\linewidth,height=5cm]{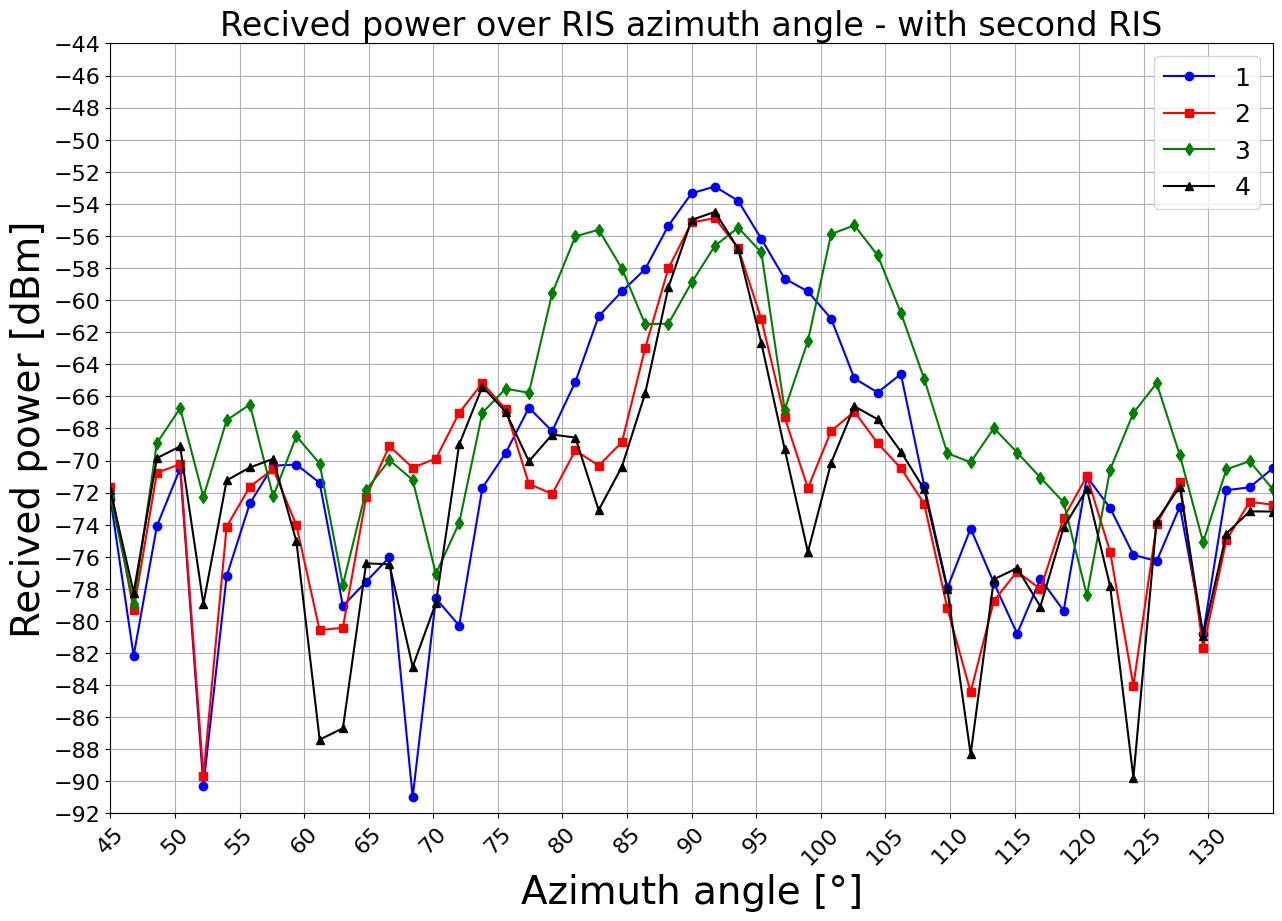}
\caption{Comparison of received power, between single RIS board and when two RIS boards are present for four selected patterns.}
\label{additional_RIS}
\end{figure}

The effect of adding a second matrix is illustrated in the plot (Fig. \ref{additional_RIS}). We observe that the inclusion of the second RIS has resulted in the appearance of many additional fades, particularly in the initial and final ranges of the azimuth angle. Notable differences in the power distribution as a function of azimuth can be observed for certain patterns. The effect was particularly visible for the first pattern. Using one matrix, three distinct peaks in received power were observed. In contrast, with two matrices, only a single wide peak was detected.  Similarly, for pattern three, there is a significant difference: with one matrix, there are five peaks, while with two matrices, there are three peaks. Overall, there is also a noticeable decrease in the received power at the transmitter. This is due to the additional non-reflective space between the antenna elements, resulting from the construction of the matrices - the PCB frame and the mounting made using 3D printing technology.

\section{Conclusion}
Presented measurement results revealed many interesting features of the RIS boards under test. The analysis of numerous 3D reflection patterns (although only four patterns have been selected for visualization in this manuscript) allows us to state that there are significant changes in the reflection characteristics depending on the selected pattern. Moreover, the second observation is that significant variations exist in the way the incident signalis reflected by the board as a function of azimuth angles. It means that for the tested RIS board, although its single antenna element is steered just by one bit, one can create a vast space of codewords (patterns); these codewords can be then used for achieving various application goals at certain locations, e.g., one should be able to choose such a 3D RIS pattern that the received signal power is maximized. The differences between the maximum and minimum value of the received power for a given pattern may be even around some tens of dB. 
However, such promising observations have to be complemented with further conclusions. One has to remember the impact of polarization; if wrongly used, the RIS boards will not provide any significant gain in terms of signal power improvement. Furthermore, the distance to the RIS has a strong impact on the observed reflection characteristics. Finally, when two RIS boards were mounted together, we observed a specific behavior of the reflection characteristics, as it became more condensed around the azimuth angles close to 90 degrees (what makes the RIS board more similar to the classic reflector).
Nevertheless, the experiments that were conducted allowed us to identify two important findings. First, one may observe significant reflection variability depending on the selected pattern; second, these patterns can be used for codebook design for some dedicated applications.

\end{document}